# The future of urban models in the Big Data and AI era: a bibliometric analysis (2000-2019).

Marion Maisonobe[1]

[1] *marion.maisonobe@cnrs.fr*
CNRS, Géographie-cités UMR 8504 CNRS, Paris (France)

**Abstract**
This article questions the effects on urban research dynamics of the Big Data and AI turn in urban management. To identify these effects, we use two complementary materials: bibliometric data and interviews. We consider two areas in urban research: one, covering the academic research dealing with transportation systems and the other, with water systems. First, we measure the evolution of AI and Big Data keywords in these two areas. Second, we measure the evolution of the share of publications published in computer science journals about urban traffic and water quality. To guide these bibliometric analyses, we rely on the content of interviews conducted with academics and higher education officials in Paris and Edinburgh at the beginning of 2018.

## 1 Introduction

Models serves in urban management for both predicting and simulating the dynamics of urban systems. However, in recent years, the "data deluge" has prompted some commentators, and in particular Chris Anderson, Editor-in-chief of Wired magazine, to question the future of models and theory (Anderson, 2008). With advances in AI and city analytics, one could similarly question the future of urban models. As discussed in the collective book "Data and the City" (Kitchin *et al.*, 2017), urban dashboards and real-time monitoring relying on a growing amount of sensors are transforming the way urban system are managed.

In this article, we show that these transformations do not necessarily imply that expert knowledge on urban systems and urban modelling methods will become obsolete. Indeed, by conducting interviews with academics in operational research, applied mathematics, hydrology, computer science and urban modelling, we found that "modelling" and "Big Data" are not necessarily exclusive and that there might be ways of hybridizing approaches.

This observation led us to formulate two hypotheses on current dynamics:
- Either academic fields traditionally related to urban management tend to improve their classic models by taking into account the growing amount of real-time data available on urban systems;
- Either the academics specialized in data handling tend to be more and more interested in the opportunity of applying AI and Big Data technologies to the management of urban systems.

Both hypotheses might be right which could result in the emergence of a new area of knowledge merging urban modelling and urban analytics or both might be wrong which would mean that so far, the transformation of urban management has no significant effect on research dynamics.

To test these hypotheses, we use two complementary materials: bibliometric data and the aforementioned interviews. We consider two areas in urban research: one, covering the academic research dealing with transportation systems and the other, with water systems. First, we measure the evolution of AI and Big Data keywords in these two areas. Second, we measure the evolution of the share of publications published in computer science journals about traffic prediction and water quality. To drive and better interpret the results of these bibliometric analyses, we rely on the content of interviews conducted with academics in Paris and Edinburgh at the beginning of 2018.

This article is organised into three sections. The first section of the article presents the state-of-the-art. The second section details the data and methodology as well as the guiding hypotheses and the third section displays the quantitative results.

## 2 The Big Data and AI era

*2.1 The Big Data and AI era in urban management*

There is a growing literature on the adoption of smart urban dashboards, on the installation of sensors and on the use of citizens' devices and cards to collect real-time data and, consequently, on the ongoing changes in the way municipalities are managing their urban systems (Bassoo *et al.*, 2018; Gray *et al.*, 2016; Jesse, 2016; Crandall, 2010). These changes are raising issues regarding the quality, volume and provenance of data (Kitchin *et al.*, 2017). The role of the citizens, their ability to take advantage and be part on the ongoing changes are also under study in the smart city literature (Komninos, 2016). Empirical research on the case of Glasgow testifies citizens have not been as much involved as it was expected so far (Borkowska and Osborne, 2018). To overcome the narrow vision of citizens as sensors (Goodchild, 2007), Borkowska and Osborne argue for making smart cities more inclusive. The social dimension of smart cities is also the focus of Anttiroiko *et al.* (2014)'s work on e-platforms. The argument is that public administrations should not adopt a narrow vision of what "smart" means by simply considering the adoption of ICTs, as the ultimate goal smart cities have to achieve. According to this branch of literature, the cities of the future should favour city dwellers' wellbeing and sustainability, and ICTs should remain a simple mean to meet these ends.

Together with this branch of literature, there is a critical literature discussing the concrete technologies and challenges at stake when referring to the buzzword "Big Data" (Chen and Zhang, 2014; Sivarajah *et al.*, 2017). Without questioning the fact that we entered the Big Data era, a collective of 21 American researchers from public and private institutions issued a white paper detailing what the analysis and practical use of large volumes of data really require (Agrawal *et al.*, 2012). This contribution also insists on the progresses Big Data has already brought in scientific fields as well as in industrial areas and urban planning. In this line, a recent article argue for the need of adapting innovative cluster policies to the challenges of the "Fourth Industrial Revolution" including Big Data's challenges (Park, 2018). On a different note, drawing on Donoho's critical views, Sha and Carotti-Sha (2016) insist more on the limits of Big Data as a "buzzword", since administrators and business circles often use it as a new and trendy word to speak about old technical issues. Together with other scholars (Kitchin *et al.* 2017), Sha and Carotti-Sha also argue that data volume matters less than the content and value of the information carried. Usefully, Batty (2016) clarifies the keyword by distinguishing between two issues: on one side, the access to new types of urban data that can change our perceptions about cities (e.g. semantic data from Social media, such as Twitter) and on the other side, traditional data that can easily become complex to handle when processed in mathematical models (e.g. Traffic data).

The debunking approach is also developing regarding "AI". Administrators and business men sometimes use "AI" as another word for "data science" although it represents just a small part of it (Carmichael and Marron, 2018) and, at the same time, it traditionally refers to a broader area of knowledge (Lungarella *et al.*, 2007). In a report, McKinsey analysts consequently distinguish between "artificial narrow AI" –potential applications of AI in business and the public sector (i.e. transfer learning, reinforcement learning and deep learning neural networks)– which is the focus of their report, and "*artificial general intelligence that could potentially perform any intellectual task a human being is capable of*" (Chui *et al.*, 2018). The relation between AI and Big Data is not straightforward, but the two concepts co-occur when referring to learning techniques. Indeed, learning techniques improve when trained on vast amounts of data. Given the growing amount of data available on urban systems, city engineers unsurprisingly consider the adoption of AI techniques for urban management (Pan *et al.*, 2016). In parallel, Pan *et al.* argue for the development of a new research area named "Urban Big Data". As we show in the next subsection, not only does existing literature question the effects of the Big Data and AI trends on the society but it sometimes simultaneously considers their effects on the academic world. Ethical and practical risks of relying on Big data and AI technologies are also under study (Kitchin, 2016; Boyd and Crawford, 2012). Together with the risks for errors when used for decision-making, there are issues regarding the use of private and personal data on individuals to administrate and monitor urban systems (Thatcher, 2014; Polenetsky and Tene, 2013). The risks for unemployment, accidents and catastrophes when used in robotics and autonomous devices (drones, vehicles) are also the topic of much research (Tuchin and Denkenberger, 2018; Stilgoe, 2017). Less studied so far is the potential for important transformations in research practices and for competitions between approaches.

The next subsection focuses on the effects of Big Data and AI on urban research, as perceived by urban research's scholars.

*2.2 The Big Data and AI era in urban research*

Anderson's provocative claim that "*the data deluge makes the scientific method obsolete*" has prompted defensive reactions in the fields of social science, human geography, urban planning and urban research (González-Bailón, 2013; Graham and Shelton, 2013; Batty, 2013; Thatcher, 2014; Bettencourt, 2014; Kitchin, 2016). For González-Bailón (2013), the increase of data in social science strengthens the need for theory. Focussing on human geography only, Graham and Shelton (2013) consider how Big Spatial Data can change the discipline of geography. Drawing upon Anderson's claim, they ask: "*What do Big Data mean for how we do research and create knowledge?*" Notwithstanding the challenge for "*conventional notions and practices of 'hard science' including to the field of Geographical Information Science*" (GIS), Graham and Shelton focus more precisely on the risk for a new quantitative and positivist turn in geography. To avoid its pitfalls, they insist on the need for critical geography and the adoption of critical views regarding Big Data. Thatcher's view (2014) is very close to theirs. He draws an interesting parallel between the advent of Spatial Big Data in the 2010s and the advent of GIS in the 1990s. According to him, a similar threat has hung over the GIS field in the 1990s, i.e. a risk for forgetting theory at the benefit of technical achievements. He recalls that, unexpectedly, GIS specialists achieved a "hard work of theory" to unravel the relation between the specific technological form and the knowledge produced. It gave birth to GIS and society, qualitative GIS, and critical GIS. According to Thatcher, a similar move toward theory is required to deal with the advent of Big Spatial Data. Addressing the issue for urban studies, Batty also rejects Anderson's claim. He particularly rejects Anderson's idea that, in the Big Data era, '*correlation supersedes causation, and science can advance even without coherent models, unified theories, or really any mechanistic explanation at all*'. According to Batty, this argument is not valid: "*In terms of cities and their functioning, the search for correlations would be something of a diversion, for what we need to look for in Big Data can only ever be discovered through the lens of theory.*"

Paradoxically, most authors seem to consider that, instead of making theory obsolete, the advent of Big Data calls for more theory. Interestingly, this claim for theory also appears in the discourse of 'hard science' specialists such as, for instance, Korudonda (2007), who advocates for more theory in the technical and applied area of Data Mining. By calling for more theory, scholars tend to favour the idea of taking into account "Big Data" as a new object of study and an opportunity in their research field. As a result, they give a description of the kind of new questions to address in the frame of social science, human geography, and urban studies. Among these scholars, Batty not only considers the challenge of Big Data for theory, but also the challenge of Big Data for the scientific practice of quantitative modelling. Drawing upon his experience in the field of urban modelling, Batty (2014) is confident that Big Data will not make urban models obsolete. The main reason for that is that urban models such as land-use models are long-term models whereas Big Urban Data is mainly about real-time data allowing for short term observations. It does not mean these data are worthless but analysing them will serve to answer new questions. As an illustration, Batty (2013) gives the example of the Oyster card data from the London subway that require the development of new models to be analysed. Regarding this last issue, Batty does not seem exactly on the same line as Bettencourt (2014) and Kitchin (2016).

According to Bettencourt (2014), modern technologies (such as deep learning techniques) can help solving difficult and important problems "*essentially without theory and this is the (potential) miracle of Big Data in cities*". However, when dealing with problems arising at longer temporal or larger scale, Bettencourt considers that the resulting complexity still require developing models of human behaviour. In this last regard, he agrees with Batty. As Bettencourt, Kitchin (2016) moderates Anderson's claim without rejecting it entirely. Instead of adopting Anderson's extreme approach to "Urban Science", Kitchin advocates for "*data-driven science that seeks to hold to the tenets of the scientific method, but seeks to generate hypotheses and insights 'born from the data' rather than 'born from the theory'*". What Bettencourt and Kitchin do not reject is that modern AI technologies and Big Data will help solving some urban and planning problems better than they were before. Considering the time and spatial scales of a range of urban issues, Bettencourt distinguishes between 1) simple issues, such as, transportation, fire, epidemics, traffic, water, trash collection, and 2) complex issues, such as, education, poverty, public housing, employment and economic development. According to him, the data-driven logic will help turning some of the first issues into simple problems and solve them.

To our view, successes of data-driven methods over theory-driven ones and scenarios of these replacements have been under-studied and under-documented so far. In this article, we address this issue through a science studies' lens. According to us, the smart city literature should not only seek to include a fourth helix, i.e. the civil society (Borkowska and Osborne, 2018), within Etzkowitz and Leydesdorff's (2000) classical model of innovation (1. government – 2. academia – 3. industry), but this literature should also put the dynamics of the academic world under renewed study. Researchers used to play an important part in the administration and monitoring of urban systems by working with public services and private companies owing the infrastructures. With the advent of Big Spatial Data owned by ICTs companies (new private players), what is the place of public researchers? Are not they threatened both by the technical challenges related to the processing of Big Data, and by the commercial nature of these data? Since Microsoft and IBM are developing their own Urban Computing teams, is there a risk for public researchers to be left behind and for public administration to be obliged to work with private researchers or to contract with new geospatial companies, such as Waze to manage their infrastructures (Courmont, 2018)? Given these novel interrogations, we consider there is a need for refocussing on the academia helix and investigate first, its internal transformation, and second, the evolution of its relations with the three other helices, namely the political world, the economic world and the social world. In this article, we want to address the first issue, academia's internal transformation in the Big Data and AI era. To do so, we conducted a set of interviews that led us, for the purpose of this article, to focus on two fields of research: the transportation field and the water field.

## 3 Interviews, guiding hypotheses and bibliometric method

### 3.1 Interviews and cases studies

To investigate the issues raised in the academic world by the advent of the Big Data and AI turn in urban management, we led two series of interviews between January and July 2018.

The first series was organised in the Eastern part of the Paris area with eleven scholars, from various applied disciplines, working in relation with a civil engineering school: three applied mathematicians, three computer scientists, two economists, one quantitative geographer, and two urban water studies' specialists. The second series was organised in Edinburgh during the launch of the Data Driven Innovation (DDI) Programme, which is part of the City Region Deal aiming to "*help establish the region as the data capital of Europe*". This programme attempts to realise its goals by helping organisations and individuals to connect to research and development in the generation, storage, analysis and use of various forms of data. In addition to attending meetings and co-organizing a round table with university officials regarding this issue, we conducted five interviews. We interviewed 1) the new centre in data science and IA's head, 2) the Data Innovation Director of the University of Edinburgh, 3) a member of the Innovation and Future Team at the City Council, 4) a co-founder of the Edinburgh Living Lab and finally, 5) a senior IT at the UBDC (Urban Big Data Center) of Glasgow University.

The interview protocol varied according to the setting. In Paris, we mainly focussed on research practices. We asked scholars their view on the possible competition between "Big Data" approaches and "modelling" approaches. Then, we asked them if they had specific examples in mind of competition between a deterministic model and a data-driven solution in any field of urban management. After that, we talked about their research practices, their possible link with "urban modelling" (in the broadest sense including all type of quantitative modelling) and with "Big Data" (in the broadest sense including all data-driven approaches). To finish, we discussed about their career. To get a comprehensive view of their position in the academic world, we invited them to comment a map of their co-authorship network. Since the interviews were semi-directed, some peripheral themes emerged and proved to be important: How to cope with the evolving needs in terms of training offers? How to deal with the gap between company and public administration's needs and research objectives? What are the true research challenges? Indeed, one of the most commonly shared idea was that most companies and public administration's needs do not represent a research challenge. According to the interviewed scholars, despite the so-called "deluge" of urban data, the scientific knowledge and the techniques necessary to meet most company and public administration's needs already exist.

As expected, these interviews highlighted some interesting cases of studies related to contemporary issues in urban systems' management. For this article, we relate two opposed cases: one case linked to traffic management and the other case linked to urban water management. The first case exemplify the possibility of hybridisation between physically inspired models and data-driven approaches and the second case, on the contrary, illustrate a case of competition between the two. We discussed the first case with two of the interviewed: an applied mathematician and a computer scientist; and we discussed the second case with two urban water studies' specialists.

The first case deals with issues of traffic simulation and real-time traffic prediction. It is the example of a PhD thesis proposing to combine stochastics methods and deterministic models (Sainct, 2016). In this research, there is a wish to improve existing models of traffic simulation that lie upon determinist laws by taking into account real-time traffic data. Pure statistical approaches based on learning algorithms are discarded in the following way: "*It should be noted that a purely statistical learning approach by time-series prediction, can in some cases give satisfactory results. On the other hand, this approach lives in a purely numerical universe. Without checking the accuracy of the sensors' physical position in relation to the reference frame, it can therefore learn an imaginary physics that has its own combinatorics and its repetitions, but has no longer any connection with reality.*" Yet, the thesis' conclusion opens with the perspective of testing neural networks to improve traffic prediction from real-time data. According to the author: "*working with real data is essential to understand traffic issues. It is through observation, and through testing, that one realizes that, although relatively frequent, the drop in capacity is a phenomenon that cannot be reproduced with the LWR model [classic model of traffic simulation].*" The author thus recognizes the weakness of classic models for real-time prediction and specifies that, to his knowledge, wishing to combine deterministic models and stochastic methods is novel and not yet conclusive. What is interesting in this example is that it led the PhD student to work both with applied mathematician specialised in traffic models and with computer scientists specialised in learning methods, and that, additionally, the student devoted part of his doctoral time on the design of a traffic monitoring application commercialised by a French multinational company. This example suggests that, at least for traffic issues, there are scholars looking for ways of combining operators' expectations, data-driven approaches and improvements of deterministic models (that try to describe the dynamics of transportation systems with physical laws). However, there might not be a majority of them and we might heard of classic models being abandoned because of their mismatch with reality. In order to get a more complete view of current dynamics in this research area, we then propose to explore its recent evolutions with bibliometric analyses.

The second case testifies from a conflict between classic modelling and real-time monitoring in the applied area of urban water research. This is the story of ProSe, a deterministic model designed to simulate the evolving water quality of a hydrographic network. Since 1995, the Siaap (a French public institution), uses this model to evaluate the impact of the Paris Region wastewater treatment system on its receiving water body: the Seine and Marne rivers (Laborie *et al.*, 2016). As explained by Laborie *et al.* (2016), the Siapp uses this model "*as an operationnal tool to assist in decision-making, such as wastewater routing choices during partial or complete WWTP stoppage.*" However, in recent years, Siaap experts and engineers began to highlight some limitations of the model regarding their actual needs. During our interviews, we discussed this specific case with water studies' specialists who conducted a fieldwork comparing this French case with a foreign one. In an article derived from their research, they focus on "*the trade-off between scientific complexity and 'usability' of scientific knowledge and tools to support management, policy and planning decisions*" (Chong *et al.*, 2017). At one point in their article, they explain that Siaap engineers are currently "*moving towards artificial intelligence and real-time control methods and are considering replacing the model with statistical techniques for daily operations (Siaap representative, 10 March 2016)."* According to our interviews with them, the reasons for this evolution are that ProSe was primarily designed for research purposes and that its designers are very keen to protect their independence and objectivity against operational needs. In addition, these scientists do not wish to include AI and real-time methods in their model because they consider these methods to relate to technical rather than scientific aspects. To solve this issue, Siaap representatives are looking for new collaborations. In particular, they intend to develop a cooperation with AI specialists located in Montréal (Canada). Even if it does not mean that the Siaap will abandon the model at the end, this story shows that there exist different views on models' role and on the scientific nature of AI and real-time control methods. Although some scholars, such as Betancourt

(2014), consider useful and promising for both science and society to apply AI and real-time control methods to urban issues, other scholars perceive these methods as outside of their research scope since they do not bring any concrete knowledge on the physical aspects of the system's dynamics. As explained by a computer scientist we interviewed, "*since about 2012, deep learning has taken a step up and made it possible to be ignorant on a subject. To have a well-functioning neural network, it only takes a little experience and the results explode the results we had before. But there is a controversy between researchers as to whether it is still science*" (Computer scientist, 16 January 2018).

From these particular cases, we note that among existing positions, some are open to the potential for hybridization between new and old methods while others see these approaches as opposed. According to the observations of Chong *et al.* (2017), it is actually common for an opposition to emerge between operational issues on the one hand, and research issues on the other. Beyond these questions of perception, there is the wariness of fad phenomena. When all funding and research policies turn to a new hot topic, schools and universities are encouraged to launch doctoral projects and new training courses on this topic in order to attract funds and students. These strategies are risky and, from the point of view of the sociologist and historian Yves Gingras, who calls for mistrust of the current enthusiasm for AI (Gingras, 2018), they can be harmful to science. During the interviews we conducted in the French school of civil engineering, researchers expressed such mistrust. These researchers remember the recent trend for financial mathematics and want to be cautious with the current craze for data science. While they consider important to develop their training offer in statistics, they do not want to transform the all curriculum according to the current interest for data-driven approaches: "*The fact that we now have access to more data than before does not justify a change in curriculum. We will certainly boost the statistics and make the student work on "reverse problems", so that they can understand when and how they can use more data in their work but that is all*" (Applied mathematician, 9 January 2018). On the other hand, at the University of Edinburgh, we observed a real willingness to develop data science courses, explained both by the university's historical strengths in AI and robotics and by the windfall effect of the data-oriented City Deal.

Without seeking to analyse the precise contents of the bibliographic sets under study, the rest of this article seeks to quantify the penetration of AI and big data methods in scientific research and production on water and transportation studies. In addition, we try to determine to what extent specialists in AI and Big Data methods tend to be interested in water and traffic issues. Indeed, it may very well happen that technique-oriented researchers end up taking up thematised subjects and proposing solutions that compete with existing solutions. In this case, these researchers may challenge traditional and expert communities or, on the contrary, they may be able to complement one another and work together. The study of the confrontation between phoneticians and acousticians during the 1970s, which led to the emergence of speech sciences analysed by Grossetti and Boë (2008), gives such an example of success of one community over the other. In the balance of this article, we conduct an exploratory analysis using bibliometric data to study the dynamics that are developing for the two particular cases we identified above: that of research on road traffic and that of research on water quality.

*3.2 The guiding hypotheses and the bibliometric approach*

Drawing upon the previous cases, we formulate two hypotheses that we propose to test on current transportation and water research dynamics:
- Either academic fields traditionally related to these research fields tend to improve their classic models by taking into account the growing amount of real-time data available on transportation and water systems;
- Either the academics specialized in data handling tend to be more and more interested in the opportunity of applying AI and Big Data technologies to the management of these two urban systems.

These hypotheses might be partly right or partly wrong since some scholars can adopt one strategy and other scholars, from the same field, can adopt a different one. The interviews led in Paris and Edinburgh suggested indeed that a variety of strategies is still prevalent at this stage. For this reason, we think that, so far, the issue addressed in this article is difficult to resolve without adopting a quantitative approach relying on bibliometrics. Even if it deprives us of the nuances accessible through interviews, it allows us to have a macroscopic view of current dynamics.

To conduct this bibliometric analysis, we propose to rely on the content of the Web of Science (WoS) database. The reason for this choice comes from the precise classification of the journals indexed in the WoS Core Collection database at the sub-discipline level. All the 10 000 journals and proceedings indexed are categorised according to one or several of the 252 distinct scientific categories among which there are "transportation" and "water resources.

In order to test the first hypothesis, we rely on these two categories. First, we select and count the publications published in the journals belonging to these two categories and second, we measure, in each set of journals, the evolving number of publications referring to AI, Big Data and other Machine Learning keywords in their titles, abstracts and list of keywords. We select AI and Big Data keywords by adapting and enriching the methodology described in publications focussing on AI publications (Cardon *et al.*, 2018) and Big Data publications (Huang *et al.*, 2015).

In order to test the second hypothesis, we select two sets of publications focussing respectively on Traffic Flow and Water Quality studies, and we measure the evolving share of the publications linked to the "Computer Science" WoS subject area compared to others WoS subject area. To select these sets, we use a selection of keywords referring to Traffic Flow and Water Quality studies. Additionally we apply lexical analyses' tools to monitor the evolution of the scientific vocabulary used in these two research areas.

If the share of AI and Big Data publications increases in traditional fields while computer science journals tend to publish a growing number of articles referring to the keywords of the two urban issues, we would consider our two hypotheses to be right. Three scenarios could be possible from this observation, among which we consider it is too soon to settle. Indeed, this situation would mean that we could equally assist in the following years to:
- The emergence of a new area of knowledge merging urban modelling and urban analytics.
- The peaceful and neutral co-existence of urban modelling and urban analytics without any merging process.
- The confrontation of urban modelling and urban analytics with one field likely to prevail over the other.

If both hypotheses were wrong, it would rather mean that so far, the transformation of urban management has no significant effect on research production dynamics in water and transportation studies. If only one of the hypothesis were right, it would mean that only urban modelling fields or only computer science fields tend to change under the influence of the transformation of urban management.

## 4 Research dynamics in urban transportation and water systems

*4.1 Research dynamics in transportation and water studies*

Transportation and water studies are both interdisciplinary research fields that are relatively autonomous from the academic disciplines they originate from. Transportation research mixes knowledge from engineering, operation research, ergonomics, computer science as well as automation and control systems whereas water research mixes environmental, engineering, ecological, meteorological and agricultural knowledge. Journals, conferences, academic departments and job positions entirely dedicated to these fields testify from their relative autonomy.

To monitor the evolving part of researches taking into account AI and Big Data approaches in these two fields, we rely on the categories available in the WoS database. In the Wos database, every journals, books and conferences are associated to at least one scientific category among 252. Some titles can belong to several categories since their content might straddle multiple categories.

At the end of February 2019, the transportation category includes 176,353 publications derived from 1441 journals, 124 books and 1358 meetings. They have been published between 1956 and 2019. On the same day, the water resources category includes 337,317 publications derived from 947 journals, 110 books and 1742 meetings. They have been published between 1961 and 2019. To monitor the evolution of AI and Big Data approaches in these two fields, we rely on two researches that propose various sets of keywords to retrieve AI and Big Data publications in the WoS database (Huang *et al.*, 2015; Cardon *et al.*, 2018). In the WoS Core Collection database, the keywords are searched in the title, the abstract and the list of keywords (both the authors' keywords and the Keyword Plus) of the publications.

Huang et al. (2015) propose five search strategies among which two lexical queries (Table 1). In Huang *et al*.'s publication, in order to reduce the noise ratio, the second lexical query is restricted by an additional set of specific keywords such as "Cloud Comput*" or "Data Min*" or "Analytic*", that need to co-occur with "Big Data" keywords. In our research, we consider this additional set of keywords is not relevant since, by searching "Big Data" publications in transportation and water resources WoS categories, we do not risk to include as much off-topic publications as if we were searching in the entire WoS Core Collection database.

| 1 | Core lexical query[1] | • TS = ("Big Data*" OR Bigdata* OR "MapReduce*" OR "Map$Reduce*" OR Hadoop* OR Hbase OR "No SQL" OR "NoSQL" OR "NoSQL Database" OR Newsql) |
|---|---|---|
| 2 | Expanded lexical query | • TS = ((Big Near/1 Data or Huge Near/1 Data) or "Massive Data" or "Data Lake" or "Massive Information" or "Huge Information" or "Big Information" or "Large-scale Data" or "Largescale Data" or Petabyte or Exabyte or Zettabyte or "Semi-Structured Data" or "Semistructured Data" or "Unstructured Data") |

**Table 1. Lexical queries used to retrieve Big Data researches (Source: Huang et al, 2015)**

To retrieve Big Data researches, we thus decide to combine the core lexical query with the first part of the expanded lexical query proposed by Huang *et al.* (Table 1). By combining these two queries, we obtain 53,086 publications published between 1960 and 2019.

Retracing the history of artificial intelligence, Cardon *et al.* (2018) propose a way to distinguish between traditional AI, also called "logic-based AI" that was dominant from the 1980s to the beginning of the 2000s, from "connexionnist AI" that is the form of AI mainly relying on artificial neural networks and deep learning techniques. Table 2 indicates the content of these two queries.

| 1 | Connexionnist AI lexical query | • TS = ("artificial neural network*" OR "Deep learning" OR "perceptron*" OR "Backprop*" OR "Deep neural network*" OR "Convolutional neural network*" OR ("CNN" AND "neural network*") OR ("LSTM" AND "neural network*") OR ("recurrent neural network*" OR ("RNN*" AND "neural network*")) OR "Boltzmann machine*" OR "hopfield network*" OR "Autoencoder*" OR "Deep belief network*") |
|---|---|---|
| 2 | Logic Based AI lexical query | • TS = ("knowledge representation*" OR "expert system*" OR "knowledge based system*" OR "inference engine*" OR "search tree*" OR "minimax" OR "tree search" OR "Logic programming" OR "theorem prover*" OR ("planning" AND "logic") OR "logic programming" OR "lisp" OR "prolog" OR "deductive database*" OR "nonmonotonic reasoning*") |

**Table 2. Lexical queries used to retrieve AI researches (Source: Cardon et al, 2018)**

To make the Connexionist AI lexical query more complete, we deliberately add the results of the following sub-query:

TS = ("adversarial neural network*" OR "generative adversarial network*" OR ("ANN$" AND "neural network*") OR ("GAN$" AND "neural network*")).

Adversarial neural networks are a class of artificial intelligence algorithms used in unsupervised machine learning and implemented by a system of two neural networks. In 2014, Goodfellow *et al*. introduced them for estimating generative models via an adversarial process. By adding the abbreviation "ANN" to the Connexionist AI lexical query, we both include publications referring to "Artificial Neural Network" and "Adversarial Neural Network", therefore improving the results of Cardon's original query. Combining Cardon's Connexionnist AI lexical query and our additional terms, we obtain 142,027 publications published between 1958 and 2019. In addition, Cardon's Logic Based AI lexical query returns 71,382 publications published between 1935 and 2019.

As we pointed out in section 2.1, AI techniques are only one subset of the techniques used to deal with vast amount of real-time data. Since there exist learning techniques that might not be considered as AI techniques, we consider a last lexical query. Drawing upon the retrieval work of Rincon-Patino *et al.*, 2018 as well as the content of the "Machine learning" Wikipedia web page[2], we come up with the following proposal:

---

[1] Here, we propose a slightly improved version of Huang *et al.*'s query since we have decided to take into account Kalantari *et al*., 2017 suggestions regarding the use of wildcards.

[2] URL: https://en.wikipedia.org/wiki/Machine_learning, retrieved the 27/02/2019

TS = ("Machine* Learn*" OR "Support Vector Machine$" OR "Support Vector Network$" OR "Random Forest$" OR "Genetic Algorithm$" OR "Bayes* Network$" OR "belief network$" OR "directed acyclic graphic*" OR "supervised learn*" OR "semi$supervised learn*" OR "unsupervised learn*" OR "reinforcement learn*" OR "turing learn*")

This last lexical query returns 313,367 publications published between 1946 and 2019. In what follows, this query will be named "Machine Learning, else" to indicate the fact that other Machine Learning publications are already taken into account in the Big Data and AI corpus of publications. Moreover, in what follows, we only consider the part of the Machine Learning set that is not included in the Big Data and AI sets. This part includes 275,770 publications. As we can observe on Figure 1, Big Data, AI (Connexionist and Logic Based) and other Machine Learning corpus of publications are only partially overlapping. One reason for this relatively small overlap can be that we only search in the titles, abstracts and keywords of the publications. However, it might also be the sign that the type of scientific contributions of each set is significantly different, which justifies treating them separately.

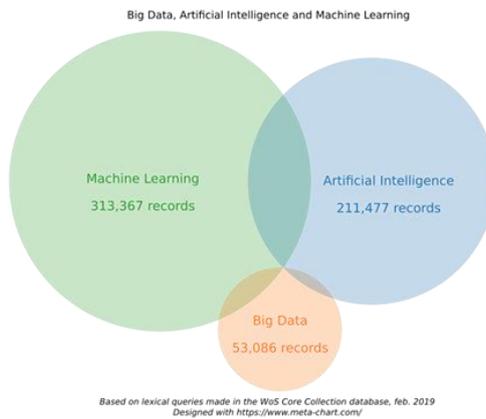

**Figure 1. Overlap between Big Data, AI and Machine Learning corpus of publication**

Figures 2 and 3 display the annual number of Big Data, Logic Based AI, Connexionist AI and Other Machine Learning publications in both the transportation and water resources WoS Categories.

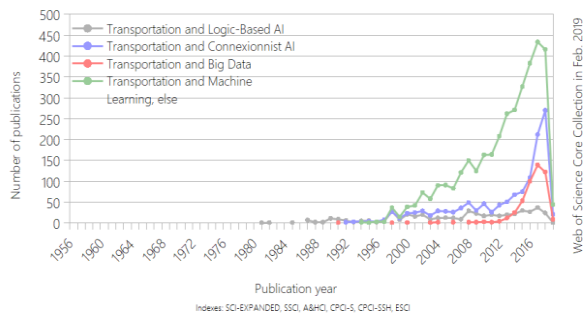

**Figure 2. Growth of Big Data, AI and other Machine Learning publications in Transportation titles**

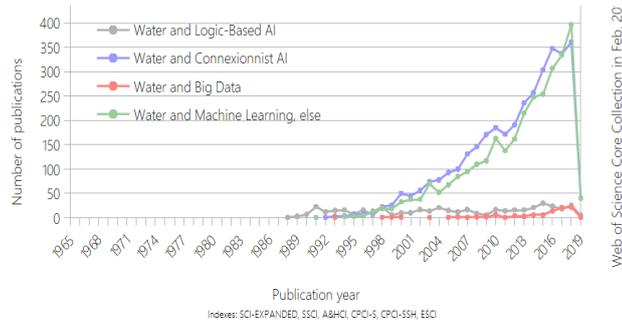

**Figure 3. Growth of Big Data, AI and other Machine Learning publications in Water Resources titles**

We observe that Machine Learning keywords tend to be increasingly used since the 2000s in both transportation and water resources' journals. Since the 2000s, Connectionist keywords have been as much used as Machine Learning keywords in water resources journals, but not in transportation journals. Connectionist keywords only developed in transportation journals after 2012 together with Big Data keywords. On the contrary, Big Data keywords do not seem to attract any interest in water resources journals. Moreover, in both transportation and water journals, Logic Based AI has never been much popular. After 2017, the number of publications is declining because the dataset is not yet complete for the years 2018 and, of course, 2019.

Drawing on Figure 2 and 3, we can confirm our first hypothesis. Indeed, these evolutions suggest that Machine Learning and AI techniques are attracting interest in both fields and that authors publishing in specialized journals in transportation and hydrology are importing and testing these techniques on their classical research problems. To verify if, conversely, transportation and water issues are attracting a growing interest among mathematicians, physicists and computer scientists specialised in Big Data and AI techniques, we propose to focus on two specific issues: traffic flow and water quality. The choice of these issues is justified by the qualitative knowledge we collected about them during the interviews' stage of our research.

*4.2 Research dynamics in traffic studies and water quality studies*

To monitor the interest for urban issues among Big Data and AI specialists, we consider the example of Traffic Flow and Water Quality studies. Drawing upon the interviews described in section 3.1, we know that both issues can be addressed using classical modelling method as well as learning techniques applied on real-time data. To select comparable set of publications associated to the two topics, we adopt a common list of action verbs "forecast", "model, "predict", "simulate" and "estimate", that we associate with keywords corresponding to the specific issues at stake. Table 3 details the resulting lexical queries that we apply for extracting WoS Core Collection publications in Traffic Flow and Water Quality studies.

| 1 | Traffic flow studies | TS=(("traffic flow*") AND (forecast* OR model* OR predict* OR simul* OR estimat*) AND (highway OR freeway OR motorway OR lane OR road OR street OR "urban network*" OR transportation)) |
|---|---|---|
| 2 | Water Quality studies | TS=((("water quality") OR (river NEAR/1 quality)) AND (forecast* OR predict* OR model* OR simul* OR estimat*) AND (phosphorus OR nitrogen OR phytoplankton)) |

**Table 3. Lexical queries used to retrieve traffic flow and water quality studies**

The resulting number of publications is similar for the two issues: 8962 publications about traffic flow and 9247 publications about water quality. In addition, the annual number of publications registered in the WoS database is similar for the two sets, which make them quite comparable (Figure 4).

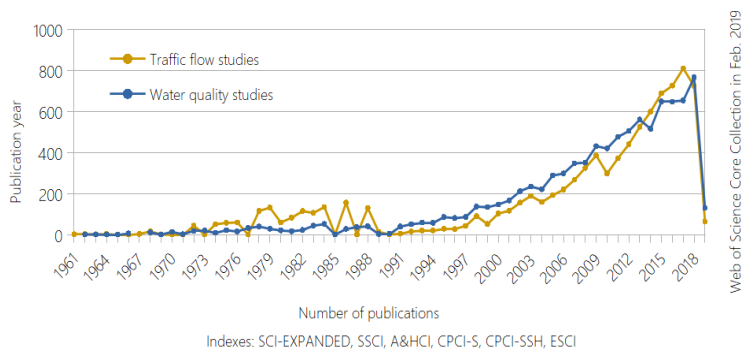

**Figure 4. Growth of traffic flow and water quality publications in the WoS Core Collection**

To monitor the dynamics of these two research issues, we adopt three strategies. First, we monitor the evolution of AI and Big Data keywords in the two sets. Second, we measure the number of traffic flow and water quality publications in computer science, mathematics and physics titles. Third, using Iramuteq software (Ratinaud, 2009), we perform an AFC analysis to detect the most significant words of the publications' abstracts before and after 2012. In addition to present the quantitative results of this bibliometric study, we use the content of our interviews to illustrate and qualify observed dynamics and discuss possible outcomes.

Big Data, AI and other Machine Learning keywords are considerably developing in Traffic Flow studies since 2005. Whereas Machine Learning keywords are much more frequent than AI and Big Data keywords in transportation journals (Figure 2), they are almost as much used as AI and Big Data keywords in Traffic Flow studies. Here, it is important to understand that the records are not limited to transportation journals. In Water Quality studies, the use of Connectionist keywords has developed around 2013 but has been declining onwards (Figure 3). Machine Learning keywords were less used than Connectionist keywords until 2018. In 2018, Machine Learning keywords were used in 27 Water Quality publications. Even, if it is more than ever before in this research area, it is still half the number of time these keywords are used in Traffic Flow studies. Corroborating the infrequent use of Big Data keywords in water resources journals, we observe Big Data keywords are almost never used in Water Quality studies. Logic-Based AI has never been much used in both Traffic Flow and Water Quality studies. Whereas there are almost the same number of Machine Learning publications in transportation and water resources journals, and there are more Connectionist publications in water resources journals than in transportation journals (Figures 2 and 3), we observe an inverse situation between Traffic Flow and Water Quality studies (Figures 5 and 6). So far, Traffic Flow is an issue leading to more AI and Machine Learning publications than Water Quality.

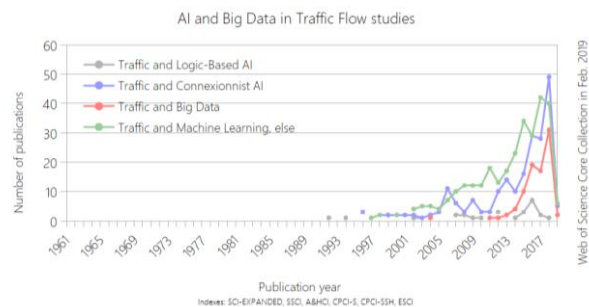

**Figure 5. Growth of Big Data, AI and other Machine Learning publications in Traffic Flow studies**

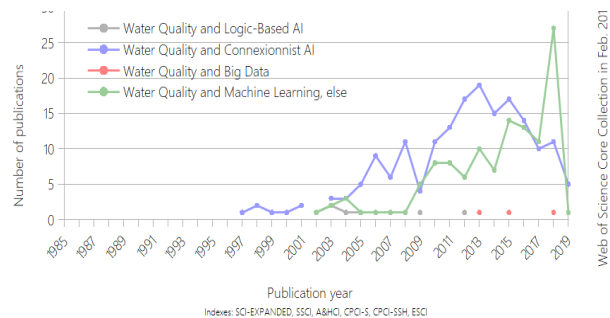

**Figure 6. Growth of Big Data, AI and other Machine Learning publications in Water Quality studies**

Traffic Flow is an issue that first made its appearance in mathematics journals in the 1960s. From there, the share of Traffic Flow publications in non-transportation journals has always been significant. On Figure 7, we observe that the number of Traffic Flow publications in journals of physics exceeded the number of Traffic Flow publications in mathematics journals from the 1990s onwards. Nagel and Schreckenberg publish their seminal contribution "A cellular automaton model for freeway traffic" in 1992 in *Journal de Physique* and, from 1991, *Physica A* issued 289 Traffic Flow publications. From the 2000s onwards, Traffic Flow publications in computer science titles started to exceed Traffic Flow publications in mathematics and physics journals. The gap is even more significant in 2009 and 2017. In 2017, the number of Traffic Flow publications in computer science journals culminates in more than 200 publications.

On the opposite, Water Quality has never been a frequent topic in mathematics, physics and computer science titles. In 2006, 2009, 2011, the topic appeared in about 20 mathematics and computer science publications but the number of Water Quality publications in these titles has continuously decreased since 2015. It suggests that, contrary to Traffic Flow, the topic does not attract professionals from modelling and prediction methods, or at least that it does not interest editorial boards of journals in mathematics, computer science and physics. Even if there is an operational interest for AI and Big Data methods to monitor water quality in urban systems, these results suggest that this interest does not translate into research dynamics. There could be various reasons for this lack of interest. Drawing on our interviews, we can think of three reasons. First, it might be that computer scientists, physicists and mathematicians have less access to water data than to traffic data. Second, it might be that the domain of validity of water quality models is bigger than traffic models' one (since vehicle's behaviour is more unpredictable than water's behaviour). Third, it might be that the water quality issue offers not enough novelty to justify a research article. As explained by a statistician we interviewed, "*in data science, problems always have different characteristics that require reflection, but this reflection is not always interesting enough to be the subject of an article in our field. You cannot make a research article for every application case.*" (Statistician, 16 January 2018).

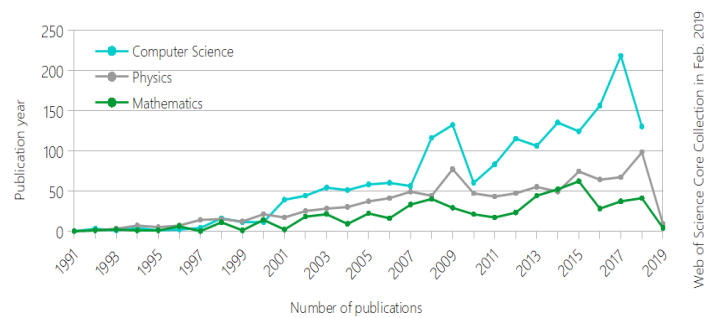

**Figure 7. Publications about Traffic Flow in computer science, mathematics and physics titles**

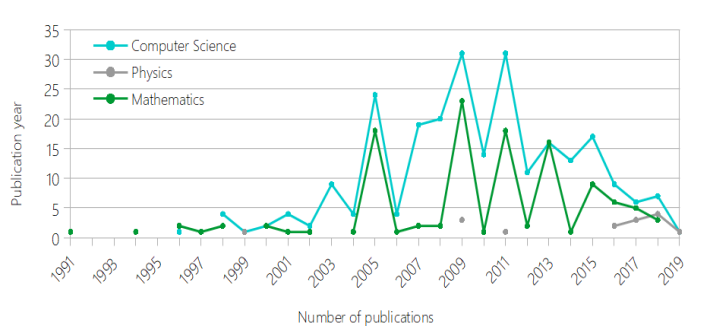

**Figure 8. Publications about Water Quality in computer science, mathematics and physics titles**

To monitor the internal dynamics of the two issues under scrutiny, we perform a lexical analysis. Considering the content of the publications' abstracts, we are able to detect, *ceteris paribus*, the most typical words used before 2012 and after 2012 for each topic (Tables 4 and 5). The choice of 2012 as time limit is justified by the fact that AI and Big Data keywords considerably developed after this date in Traffic Flow studies (Figure 5). Therefore, we suspect the vocabulary used in Traffic Flow studies would significantly reflect the growing adoption of AI and Big Data techniques after 2012.

As expected, we notice that the word "deep" is one of the most significantly used words after 2012 in Traffic Flow studies contrary to the nominal group "cellular automaton". It suggests, in line with the assumptions of this work, that the physically inspired "cellular automaton" method is in the process of being replaced by "deep learning" methods to monitor and predict traffic flow. The word "model" is also typical of pre-2012 Traffic Flow studies, which suggest that, at least for traffic issues, we might assist to "the end of theory" announced by Anderson. However, if we look to other post-2012 words, we observe the acronym "MFD" together with an interest for new types of vehicles (autonomous vehicles, smart vehicles and bicycles). It suggest that, in addition to a novel consideration for smart mobility and bicycles, Traffic Flow studies are still focussing on the fundamental diagram. This diagram gives a relation between traffic flux and traffic density. Traffic operators use it to monitor urban congestion. In line with the example developed in 3.1, it seems that, so far, Traffic Flow research is integrating AI and Big Data techniques without abandoning classical approaches. To confirm this observation, we look at the evolution of the number of Traffic Flow publications in two important transportation journals: Transportation Research Part B and Transportation Research Part C. The difference between the two journals is topical: Part B focusses on physical models whereas Part C focusses on new technologies. As shown in Figure 9, the number of traffic flow publications published in Part C increased in recent years, but the issue remained important in Part B. According to us, monitoring the evolving distribution of Traffic Flow publications between these two journals is a good way to track the research dynamics of the research area. Taking into account the evolving citation behaviour of the authors publishing in these two journals could also be an interesting way of following the current dynamics (analysing the evolving scope of their cited references).

| Typical words before 2012 | Concentration; canyon; transition; cellular automaton; phase; highway; model; ramp; freeway; flow; incident; AHS (automated highway system); pollutant; computer |
|---|---|
| Typical words from 2012 onwards | Connect; propose; VSL (variable speed limit); vehicle; prediction; energy; bicycle; autonomous; smart; electric; deep; consumption; mobility; MFD (macroscopic fundamental diagram); cooperative |

**Table 4. Typical words used before and after 2012 in Traffic Flow studies**

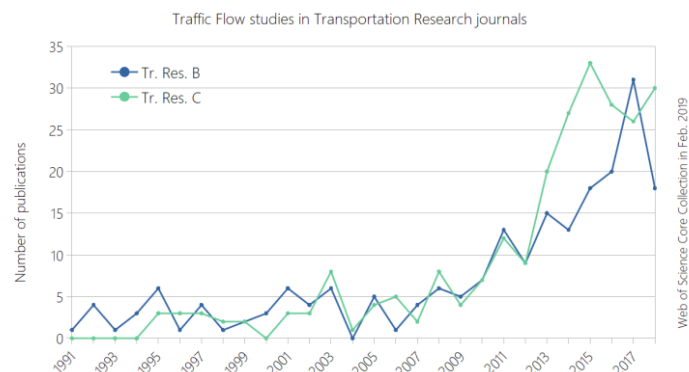

**Figure 9. Traffic Flow studies in Transportation Research journals**

As we could expect drawing on Figures 6 and 8, the lexical evolution observed in Water Quality studies does not allow identifying a change in research methods. The main difference seems to be a geographic difference since the words "China" and "TGR" (Three Gorges Reservoir) make their appearance after 2012. It suggest that Water Quality research has become an important issue in China in recent years. The word "scenario" which is a feature of modelling approach is one on the most typical words used after 2012, suggesting that learning techniques do not threaten modelling approaches in this research area. Once again, given the observations made in interviews, we can make the hypothesis that, so far, while learning techniques interest water operators, they do not modify research dynamics among Water Quality scholars. Figure 2 confirms that AI techniques are known and used for a long time in water research, but according to Figures 6 and 8, Water Quality does not appear to be an interesting issue for applying them, at least for research purpose.

| Typical words before 2012 | Manure; soil; ground; plot; NH (Ammonia Nitrogen); lagoon; litter; runoff; mug; application; GIS; waste; poultry; deposition; compost |
|---|---|
| Typical words from 2012 onwards | Climate; China; TN (Total Nitrogen); driver; service; change; study; legacy; future; TGR (Three Gorges Reservoir); scenario; switchgrass; mitigation; Chl (Chlorophyll); stormwater |

**Table 5. Typical words used before and after 2012 in Water Quality studies**

Given the results of this bibliometric analysis, we can establish that the second hypothesis is true for Traffic Flow studies, but wrong for Water Quality studies. In particular, we observe a growing interest for AI and Big Data technologies applied to traffic flow data in Computer Science, but not to water quality data.

## 5 Discussion

This exploratory work suggests that the effects of the Big Data and AI turn in urban management vary according to urban issues. Focussing on Traffic Flow studies, we observed that AI, Big Data and

Machine Learning keywords are increasingly used and that computer science titles are publishing a growing number of publications on the topic. On the opposite, computer science, mathematics and physics titles tend to publish little on water quality. Moreover, the use of AI and Machine Learning keywords is not growing as much in Water Quality studies as in Traffic Flow studies. As shown by the interviews, scholars adopt various strategies regarding the fad for AI and Machine Learning methods. Certain scholars see these methods as not useful for their research purposes while other consider them as an opportunity for improving classical models and offer better solutions to operational needs. In the vision area, when enough data are available, deep learning methods are now considered more efficient than any other supervised method used in the past, but in most fields, the need for modelling seems far from over. According to one of our interviewees: *"There are still some grey areas, situations on which we do not have access to the data and in these cases, we still need to simulate them. There is also the problem of exploiting raw data. In general, it is always useful to pre-process the data and for this purpose, the accumulated knowledge can be useful." (Applied mathematician, 9 January 2018).* In addition to the need of enriching the data, it is also important to consider the research purpose. Learning methods might not be suitable for solving all urban problems, and currently, they are mostly used for prediction purposes. Yet, according to the statistician we interviewed, there is a growing interest for optimisation and predictive maintenance issues in Machine Learning and these perspectives might require more interdisciplinary in the future. Finally, maybe one of the main pitfalls of machine learning techniques applied to big amounts of data so far is the one cited by one of the economists we met: *"With Big Data, the computer will find patterns and laws of evolution but it only works if there really are regularities, so it works if the regularities are maintained." (Economist, 17 February 2018)*. Working on LUTI models (land-use-transport-interaction), this economist is interested in developing operational tools for planning in the long term. In line with what Batty observed regarding mobility data (2013) and with Te Brömmelstroet *et al.*, (2014), learning methods are not ready to replace classic modelling for such purposes. Hybridising approaches might be an interesting avenue as exemplified by Traffic Flow studies (sections 3.1 and 4.2), but as objected by one of our interviewees, hybridising approaches is challenging and does not always bring exploitable results (Computer scientist, 8 February 2018).

## 6 Conclusion

In this article, we monitored the adoption of AI and Big Data approaches in two urban fields: transportation and water. To do so, we relied on the content of interviews and on bibliometric data. In particular, we measured the evolution of the number of publications using AI and Big Data keywords in the two fields under study. To detect if these urban issues were also attracting interest among specialists of AI and Big Data methods, we measured the number of publications on traffic flow and water quality in computer science, mathematics and physics journal. To finish, we focussed on significant changes in the vocabulary used in traffic flow and water quality publications' abstracts before and after 2012. This method helped us identifying significant differences between the two urban issues. Traffic Flow studies are focussing more and more on AI and Big Data and are arousing a growing interest among computer scientists while Water Quality studies do not. To go further in the understanding of current research dynamics, we think that other approaches could be interesting. In particular, studying the evolving citation behaviours of authors publishing in urban fields might help identifying ongoing transformations within these fields. In addition, it could be useful to follow the publication behaviour of a specific cohort of scholars specialized in a given urban issue. Doing so could help distinguishing between scholars that consider new techniques by citing external publications on AI and Big Data and scholars that keep referring to traditional references of their field and applying classical modelling methods. In addition, it could be interesting to focus on the locations and institutional affiliations of these authors. Indeed, we could imagine that scholars' strategies varies according to their setting. In the case of the two places where we conducted interviews, we observed very distinct discourses driven by very distinct institutional strategies.

**Acknowledgments**

**Appendix**

Lexical queries used to retrieve Big Data, AI and Machine Learning publications

| Logic Based AI | TS = ("knowledge representation*" OR "expert system*" OR "knowledge based system*" OR "inference engine*" OR "search tree*" OR "minimax" OR "tree search" OR "Logic programming" OR "theorem prover*" OR ("planning" AND "logic") OR "logic programming" OR "lisp" OR "prolog" OR "deductive database*" OR "nonmonotonic reasoning*") |
|---|---|
| Connectionist AI | TS = ("artificial neural network*" OR "Deep learning" OR "perceptron*" OR "Backprop*" OR "Deep neural network*" OR "Convolutional neural network*" OR ("CNN" AND "neural network*") OR ("LSTM" AND "neural network*") OR ("recurrent neural network*" OR ("RNN*" AND "neural network*")) OR "Boltzmann machine*" OR "hopfield network*" OR "Autoencoder*" OR "Deep belief network*" OR "adversarial neural network*" OR "generative adversarial network*" OR ("ANN$" AND "neural network*") OR ("GAN$" AND "neural network*")) |
| Big Data | TS = ("Big Data*" OR Bigdata* OR "MapReduce*" OR "Map$Reduce*" OR Hadoop* OR Hbase OR "No SQL" OR "NoSQL" OR "NoSQL Database" OR Newsql OR Big Near/1 Data or Huge Near/1 Data) OR "Massive Data" OR "Data Lake" OR "Massive Information" OR "Huge Information" OR "Big Information" OR "Large-scale Data" OR "Largescale Data" OR Petabyte OR Exabyte OR Zettabyte OR "Semi-Structured Data" OR "Semistructured Data" OR "Unstructured Data") |
| Machine Learning, else | TS = ("Machine* Learn*" OR "Support Vector Machine$" OR "Support Vector Network$" OR "Random Forest$" OR "Genetic Algorithm$" OR "Bayes* Network$" OR "belief network$" OR "directed acyclic graphic*" OR "supervised learn*" OR "semi$supervised learn*" OR "unsupervised learn*" OR "reinforcement learn*" OR "turing learn*") NOT TS = ("knowledge representation*" OR "expert system*" OR "knowledge based system*" OR "inference engine*" OR "search tree*" OR "minimax" OR "tree search" OR "Logic programming" OR "theorem prover*" OR ("planning" AND "logic") OR "logic programming" OR "lisp" OR "prolog" OR "deductive database*" OR "nonmonotonic reasoning*" OR "artificial neural network*" OR "Deep learning" OR "perceptron*" OR "Backprop*" OR "Deep neural network*" OR "Convolutional neural network*" OR ("CNN" AND "neural network*") OR ("LSTM" AND "neural network*") OR ("recurrent neural network*" OR ("RNN*" AND "neural network*")) OR "Boltzmann machine*" OR "hopfield network*" OR "Autoencoder*" OR "Deep belief network*" OR "adversarial neural network*" OR "generative adversarial network*" OR ("ANN$" AND "neural network*") OR ("GAN$" AND "neural network*") OR "Big Data*" OR Bigdata* OR "MapReduce*" OR "Map$Reduce*" OR Hadoop* OR Hbase OR "No SQL" OR "NoSQL" OR "NoSQL Database" OR Newsql OR Big Near/1 Data or Huge Near/1 Data) OR "Massive Data" OR "Data Lake" OR "Massive Information" OR "Huge Information" OR "Big Information" OR "Large-scale Data" OR "Largescale Data" OR Petabyte OR Exabyte OR Zettabyte OR "Semi-Structured Data" OR "Semistructured Data" OR "Unstructured Data") |